\documentclass[final,3p,times]{elsarticle}

\usepackage{amsmath,amssymb,mathtools}
\usepackage{booktabs}
\usepackage{siunitx}
\usepackage{graphicx}
\usepackage[dvipsnames]{xcolor}
\usepackage{caption}
\usepackage{subcaption}
\usepackage{multirow}
\usepackage{placeins}   
\usepackage{microtype}
\usepackage{enumitem}
\usepackage[hidelinks]{hyperref}
\usepackage[T1]{fontenc}
\usepackage[utf8]{inputenc}

\DeclareSIUnit{\kbps}{kbps}

\graphicspath{{figures/}} 

\biboptions{sort&compress}

\journal{}

\begin{document}

\begin{frontmatter}

\title{ISO/IEC-Compliant Match-on-Card Face Verification with Short Binary Templates}

\author[uiz]{Abdelilah Ganmati\corref{cor1}}
\author[uiz]{Karim Afdel}
\author[uiz]{Lahcen Koutti}
\cortext[cor1]{Corresponding author.}
\ead{a.ganmati@uiz.ac.ma}
\ead[kaf]{k.afdel@uiz.ac.ma}
\ead[lk]{l.koutti@uiz.ac.ma}

\address[uiz]{Computer Systems \& Vision Laboratory, Faculty of Sciences, Ibn Zohr University, BP 8106, Agadir 80000, Morocco}

\begin{abstract}
We present a practical match-on-card design for face verification in which compact 64/128-bit templates are produced off-card by PCA--ITQ and compared on-card via constant-time Hamming distance. We specify ISO/IEC~7816-4/14443-4 command APDUs with fixed-length payloads and decision-only status words (no score leakage), together with a minimal per-identity EEPROM map. Using real codes from a CelebA working set (55 identities, 412 images), we (i) derive operating thresholds from ROC/DET, (ii) replay enrol$\rightarrow$verify transactions at those thresholds, and (iii) bound end-to-end time by pure link latency plus a small constant on-card budget. Even at the slowest contact rate (9.6\,kbps), total verification time is 43.9\,ms (64\,b) and 52.3\,ms (128\,b); at 38.4\,kbps both are $<14$\,ms. At FAR$=1\%$, both code lengths reach TPR$=0.836$, while 128\,b lowers EER relative to 64\,b. An optional $+6$\,B helper (symbol-level parity over empirically unstable bits) is latency-negligible. Overall, short binary templates, fixed-payload decision-only APDUs, and constant-time matching satisfy ISO/IEC transport constraints with wide timing margin and align with ISO/IEC~24745 privacy goals. Limitations: single-dataset evaluation and design-level (pre-hardware) timing; we outline AgeDB/CFP-FP and on-card microbenchmarks as next steps.
\end{abstract}


\end{frontmatter}

\section{Introduction}
Short, fixed-length binary face templates enable constant-time Hamming matching and tiny on-card storage—an excellent fit for contact and contactless smart cards. Classical hashing methods such as Spectral Hashing (SH) and Iterative Quantization (ITQ) compress high-dimensional embeddings into compact codes while preserving neighborhood structure; in practice, a PCA front-end plus ITQ rotation gives strong binarization quality at very low bit budgets \cite{Weiss2008SpectralHashing,Gong2011ITQ}. With modern identity-supervised backbones (e.g., MobileFaceNets ) providing highly discriminative float features, 64–128\,bit PCA–ITQ templates commonly support millisecond-scale comparisons and minimal APDU payloads \cite{Chen2018MobileFaceNets}.

\paragraph{Challenge.}
Deploying \emph{match-on-card} in real systems hinges on three constraints: (i) conformance to ISO/IEC 7816-4 (contact) and ISO/IEC 14443-4 (contactless) transport behavior—fixed payload sizes, status words, and robust chaining; (ii) a lean memory map and constant-time matching on the Secure Element (SE); and (iii) a privacy posture aligned with ISO/IEC 24745 (non-invertibility, revocability, unlinkability) \cite{ISO7816,ISO14443,ISO24745}. Beyond protocol compliance, developers need transaction-level behavior that agrees with ROC-derived operating thresholds on \emph{real} binary templates, and end-to-end latency that is bounded primarily by link speed rather than on-card computation.

\paragraph{Contributions.}
\begin{itemize}[leftmargin=1.05em]
\item \textbf{ISO/IEC-compliant design and interfaces.} We specify fixed-length \textsf{ENROLL\_TEMPLATE} and \textsf{VERIFY\_BINARY} APDUs with decision-only returns (no score leakage), constant-time XOR+\texttt{popcount} on-card, and strict \texttt{Lc}/\texttt{Le} checking, directly mapping to ISO/IEC~7816-4/14443-4. 
\item \textbf{Operating points on real codes.} On a CelebA working set (55 identities, 412 images), two-image majority enrollment yields $\mathrm{TPR}@\mathrm{FAR}{=}1\%=0.836$ for both 64\,b and 128\,b; 128\,b achieves a lower EER at the cost of +8\,B storage/wire.
\item \textbf{Transport-bounded latency.} Simulations across ISO/IEC~7816-4 (9.6/38.4/115.2\,kbps) and ISO/IEC~14443-4 (106–848\,kbps) show all configurations $\ll\!100$\,ms end-to-end; at 38.4\,kbps both 64\,b and 128\,b complete in $<14$\,ms.
\item \textbf{Privacy posture.} Decision-only status words avoid score leakage; PCA decorrelation plus ITQ rotation followed by sign binarization hinder inversion; revocability/diversity are supported via per-application RotationIDs, aligning with ISO/IEC~24745 requirements.
\end{itemize}

\section{Related Work}
\textbf{Binary hashing for compact descriptors.}
Classical unsupervised hashing methods—Spectral Hashing (SH) and Iterative Quantization (ITQ)—map high-dimensional descriptors to short binary codes while preserving neighborhood structure; PCA followed by an ITQ rotation is a strong, simple baseline that performs well at very low bit budgets \cite{Weiss2008SpectralHashing,Gong2011ITQ}. Deep variants improve compactness and robustness, e.g., DeepBit (unsupervised) and deep supervised discrete hashing (DSH), which directly optimize binary outputs using label information \cite{Lin2016DeepBit,Liu2016DSH}. For face verification, identity-supervised embeddings such as MobileFaceNets provide highly separable float spaces that remain discriminative after PCA–ITQ binarization \cite{Chen2018MobileFaceNets}.

\textbf{Template protection paradigms.}
Two families dominate: \emph{biometric cryptosystems} using helper data/Error-Correcting Codes (e.g., fuzzy commitment, fuzzy extractors) \cite{Juels1999FuzzyCommitment,Dodis2008FuzzyExtractors}, and \emph{cancelable biometrics} (non-invertible, revocable transforms) \cite{Ratha2001Cancelable}. Surveys highlight the security–accuracy trade-offs and the difficulty of achieving strong irreversibility and unlinkability without sacrificing recognition \cite{Jain2008TemplateSecurity}. More recent work formalizes ISO/IEC 24745 criteria (irreversibility, unlinkability, renewability) and proposes practical evaluations and unlinkability metrics \cite{GomezBarrero2016UnlinkableISO}.
Targeted redundancy for unstable bits in binary face templates was recently introduced in \cite{Ganmati2025TRS}; here we quantify its transport overhead within ISO/IEC~7816/14443 and show it is latency-negligible.

\textbf{Decision interfaces and privacy leakage.}
A well-known attack surface in deployed systems is score leakage (hill-climbing and inference from similarity scores). Architectures that \emph{return only a decision}—no scores—reduce this surface and align with privacy-by-design principles seen in privacy-preserving biometric protocols and encrypted matching \cite{Bringer2009EncryptedID}. Our design follows this line: strict fixed-length APDUs and decision-only status words.

\textbf{Standards and transport.}
Interoperable smart-card transport is governed by ISO/IEC 7816-4 (contact) and ISO/IEC 14443-4 (contactless); biometric information protection goals are set by ISO/IEC 24745. Our interface choices (fixed \texttt{Lc}/\texttt{Le}, constant-time Hamming, decision-only returns, revocability handles) are made to conform with these documents \cite{ISO7816,ISO14443,ISO24745}.

\section{System Overview}

\subsection{Binary template generation (off-card)}
Given a float embedding $\mathbf{f}\in\mathbb{R}^{d}$ from a supervised face encoder, we form a compact binary template via a linear PCA front-end followed by an ITQ rotation and a componentwise sign binarizer:
\begin{equation}
\mathbf{x} = (\mathbf{f}-\boldsymbol{\mu})\,W_{\mathrm{PCA}},\qquad
\mathbf{z} = \mathbf{x}\,R,\qquad
\mathbf{b} = \mathrm{sign}(\mathbf{z}) \in \{0,1\}^{L},\quad L\in\{64,128\},
\label{eq:pcaitq}
\end{equation}
where $W_{\mathrm{PCA}}\in\mathbb{R}^{d\times L}$ keeps the top-$L$ principal directions and $R\in\mathbb{R}^{L\times L}$ is the orthogonal ITQ rotation \cite{Gong2011ITQ}. We define $\mathrm{sign}(t)=\mathbb{1}[t>0]$ to produce bits directly in $\{0,1\}$. 
For revocability and application diversity, deployments maintain per-application \emph{RotationID} handles that select $(W_{\mathrm{PCA}},R)$ (and optional salts) from a versioned registry; changing the RotationID triggers re-enrollment to yield an unlinkable template under the new parameters.

\paragraph{Bit packing and transport.}
Binary templates are serialized MSB-first into $L/8$ bytes (2, 4, 8, or 16 bytes). This byte string is used verbatim in APDU payloads for enrollment and verification.\footnote{We use big-endian multi-byte integers and MSB-first bit packing for interoperability with ISO/IEC 7816-4/14443-4 tooling;} 

\subsection{On-card matching (constant time)}
For each identity, the Secure Element (SE) stores a single enrolled template $\mathbf{b}_{\mathrm{enr}}$ in protected EEPROM together with minimal metadata (RotationID, policy flags). At verification, the host sends a probe template $\mathbf{b}_{\mathrm{prb}}$; the card computes the Hamming distance
\[
D_H(\mathbf{b}_{\mathrm{prb}},\mathbf{b}_{\mathrm{enr}})=\mathrm{popcount}\!\left(\mathbf{b}_{\mathrm{prb}}\oplus\mathbf{b}_{\mathrm{enr}}\right),
\]
using a fixed loop over $L/8$ bytes (no data-dependent branches). The threshold test $D_H\le \tau$ is implemented with constant-time comparisons and returns \emph{only} a decision via status words (e.g., \texttt{0x9000} accept, \texttt{0x6985} reject). No similarity scores or intermediate values are ever returned.

\paragraph{Memory and timing footprint.}
The per-identity footprint is $L/8$ bytes for the template plus 2–4 bytes of RotationID and 1 byte of policy flags (integrity tags are optional. XOR+\texttt{popcount} executes in sub-millisecond time for $L\!\in\!\{64,128\}$; in our transport-bounded measurements the end-to-end time is dominated by the ISO/IEC link, not on-card compute.

\paragraph{Revocation and renewal.}
A dedicated \textsf{REKEY\_ROTATION} command updates RotationID (and optional SaltID). Applications then re-issue fresh templates; the old reference is overwritten. This enables revocation/renewal without exposing raw biometrics and supports per-application diversity.

\begin{figure}[!t]
  \centering
  \includegraphics[width=0.70\linewidth]{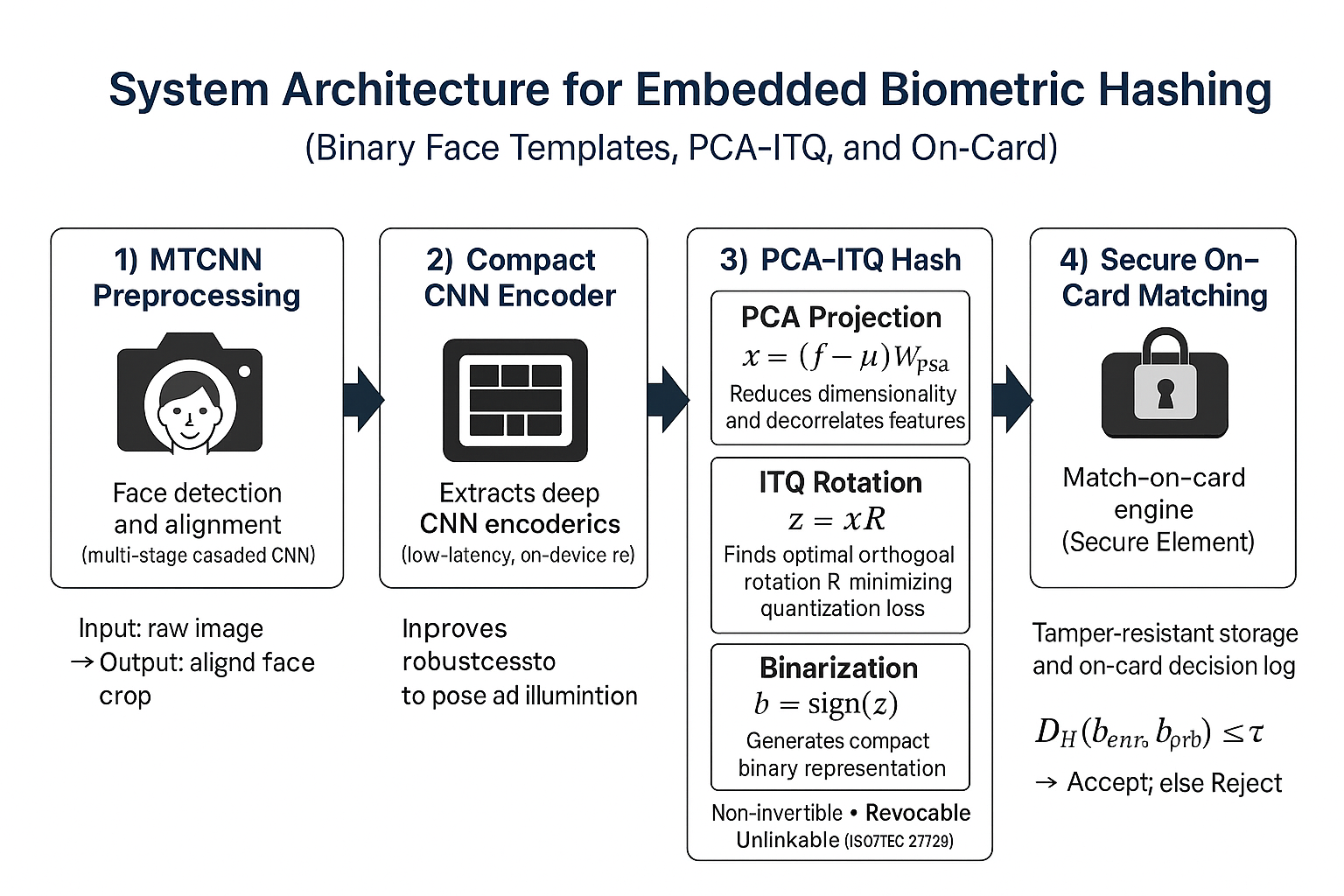}
  \caption{End-to-end pipeline. Off-card: detection/alignment, backbone encoder, PCA--ITQ hashing to $L\!\in\!\{64,128\}$ bits. On-card: constant-time XOR+\texttt{popcount}, decision-only APDUs. Transport uses ISO/IEC~7816-4 (contact) and ISO/IEC~14443-4 (contactless).}
  \label{fig:arch}
\end{figure}
\FloatBarrier

\subsection{Threat model and compliance posture}
\label{subsec:threat}
We assume an honest-but-curious host and a physically tamper-resistant SE. The adversary may observe APDU traffic and attempt hill-climbing or inference from responses. Our design aligns with ISO/IEC~24745 as follows: (i) \emph{Non-invertibility}—templates are linear-projected and rotated before sign binarization; only compact bits are stored on-card, not floats. (ii) \emph{Unlinkability}—the interface returns decisions only (no scores), and RotationIDs/salts yield application-specific templates. (iii) \emph{Revocability}—RotationID changes force re-enrollment to fresh, unlinkable references. Secure messaging (e.g., platform-provided SCP) can be enabled to protect APDU payloads when confidentiality/integrity on the transport is required; it is orthogonal to template protection and does not alter payload formats .

\subsection{Engineering notes (on-card hygiene)}
\label{subsec:eng-notes}
To reduce side channels and state leakage: (i) implement fixed-bound loops and constant-time comparisons; (ii) clear transient probe buffers before return; (iii) enforce strict \texttt{Lc}/\texttt{Le} and header checks; (iv) optionally rate-limit repeated \textsf{VERIFY\_BINARY} attempts and maintain lockout counters as policy.

\section{APDUs, Memory Map, and Policies}
We use \emph{fixed-length} payloads, strict \texttt{Lc}/\texttt{Le} checking, and \emph{decision-only} status words. Malformed length $\Rightarrow$ \texttt{0x6700}; unknown \texttt{INS} $\Rightarrow$ \texttt{0x6D00}; wrong data/format $\Rightarrow$ \texttt{0x6A80}. Table~\ref{tab:apdus} summarizes the interface; Table~\ref{tab:memory} gives the EEPROM footprint. All multi-byte integers are big-endian, and bitstrings are packed MSB-first.

\begin{table}[!t]
\centering
\caption{APDU synopsis (implementation-neutral fields; short APDUs suffice for $L\le 128$).}
\label{tab:apdus}
\small
\begin{tabular}{@{}llp{9.1cm}@{}}
\toprule
\textbf{Command} & \textbf{CLA/INS} & \textbf{Data payload (bytes)} \\ \midrule
\textsf{ENROLL\_TEMPLATE} & 0x80 / 0x10 &
Version(1), HashLenBits(1), RotationID(2), $\mathbf{b}_{\text{enr}}$ ($L/8$) \\
\textsf{VERIFY\_BINARY}   & 0x80 / 0x20 &
Version(1), HashLenBits(1), RotationID(2), $\mathbf{b}_{\text{prb}}$ ($L/8$) \\
\textsf{REKEY\_ROTATION}  & 0x80 / 0x30 &
NewRotationID(2) \\ \bottomrule
\end{tabular}
\end{table}

\begin{table}[!t]
\centering
\caption{Per-identity EEPROM footprint (without/with optional integrity tag).}
\label{tab:memory}
\small
\begin{tabular}{@{}lcccc@{}}
\toprule
\textbf{Bits} & Template (B) & RotationID (B) & Policy (B) & Total (no MAC \;/\; +8..16 B) \\
\midrule
16  & 2  & 2 & 1 & 5  \;/\; 13--21 \\
32  & 4  & 2 & 1 & 7  \;/\; 15--23 \\
64  & 8  & 2 & 1 & 11 \;/\; 19--27 \\
128 & 16 & 2 & 1 & 19 \;/\; 27--35 \\
\bottomrule
\end{tabular}
\end{table}

\noindent\textbf{Status words (decision-only policy).}
\begin{table}[!t]
\centering
\label{tab:sws}
\small
\begin{tabular}{@{}ll@{}}
\toprule
\textbf{SW} & \textbf{Meaning} \\
\midrule
\texttt{0x9000} & Success (OK). For \textsf{VERIFY\_BINARY}: \emph{accept} ($D_H \le \tau$). \\
\texttt{0x6985} & Conditions not satisfied. For \textsf{VERIFY\_BINARY}: \emph{reject} ($D_H > \tau$). \\
\texttt{0x6A80} & Wrong data (unsupported \texttt{HashLenBits}, inconsistent headers, malformed payload). \\
\texttt{0x6A82} & Record not found (e.g., unknown \texttt{TemplateID} if enabled). \\
\texttt{0x6A84} & Not enough memory space (EEPROM quota exceeded). \\
\texttt{0x6982} & Security status not satisfied (issuer/auth not established). \\
\texttt{0x6700} & Wrong length (\texttt{Lc}/\texttt{Le} mismatch, non\-canonical size). \\
\texttt{0x6D00} & Instruction code not supported (unknown \texttt{INS}). \\
\bottomrule
\end{tabular}
\caption*{Decision-only interface: no similarity scores are returned.}
\end{table}

\noindent\textbf{Byte-level layouts (MSB-first packing).} Offsets are relative to the start of the APDU data field; fields in \emph{italics} are optional and disabled in our core experiments.

\begin{table}[!t]
\centering
\small
\caption{\textsf{ENROLL\_TEMPLATE}: data payload layout (host$\rightarrow$card).}
\label{tab:enroll-layout}
\begin{tabular}{@{}ccccp{8.9cm}@{}}
\toprule
\textbf{Offset} & \textbf{Len} & \textbf{Type} & \textbf{Name} & \textbf{Notes} \\
\midrule
0 & 1 & u8  & Version        & Protocol version (e.g., 0x01). \\
1 & 1 & u8  & HashLenBits    & \textbf{16, 32, 64, 128}. \\
2 & 2 & u16 & RotationID     & Selects $(W_{\mathrm{PCA}},R)$ on card. \\
4 & \emph{2} & \emph{u16} & \emph{SaltID}       & \emph{Per-application diversity; default unset.} \\
6 & \emph{2} & \emph{u16} & \emph{TemplateID}   & \emph{Record selector if multiple templates.} \\
8 & $L/8$ & bytes & $\mathbf{b}_{\text{enr}}$ & Binary template, MSB-first within each byte. \\
\bottomrule
\end{tabular}
\end{table}

\begin{table}[!t]
\centering
\small
\caption{\textsf{VERIFY\_BINARY}: data payload layout (host$\rightarrow$card).}
\label{tab:verify-layout}
\begin{tabular}{@{}ccccp{8.9cm}@{}}
\toprule
\textbf{Offset} & \textbf{Len} & \textbf{Type} & \textbf{Name} & \textbf{Notes} \\
\midrule
0 & 1 & u8  & Version        & Must match supported protocol version. \\
1 & 1 & u8  & HashLenBits    & \textbf{16, 32, 64, 128}. \\
2 & 2 & u16 & RotationID     & Must match enrolled parameters. \\
4 & \emph{2} & \emph{u16} & \emph{SaltID}       & \emph{If set at enrollment.} \\
6 & \emph{2} & \emph{u16} & \emph{TemplateID}   & \emph{If multiple references exist.} \\
8 & $L/8$ & bytes & $\mathbf{b}_{\text{prb}}$ & Probe template, MSB-first. \\
\bottomrule
\end{tabular}
\end{table}

\FloatBarrier
\begin{figure}[!t]
  \centering
  \includegraphics[width=0.42\linewidth]{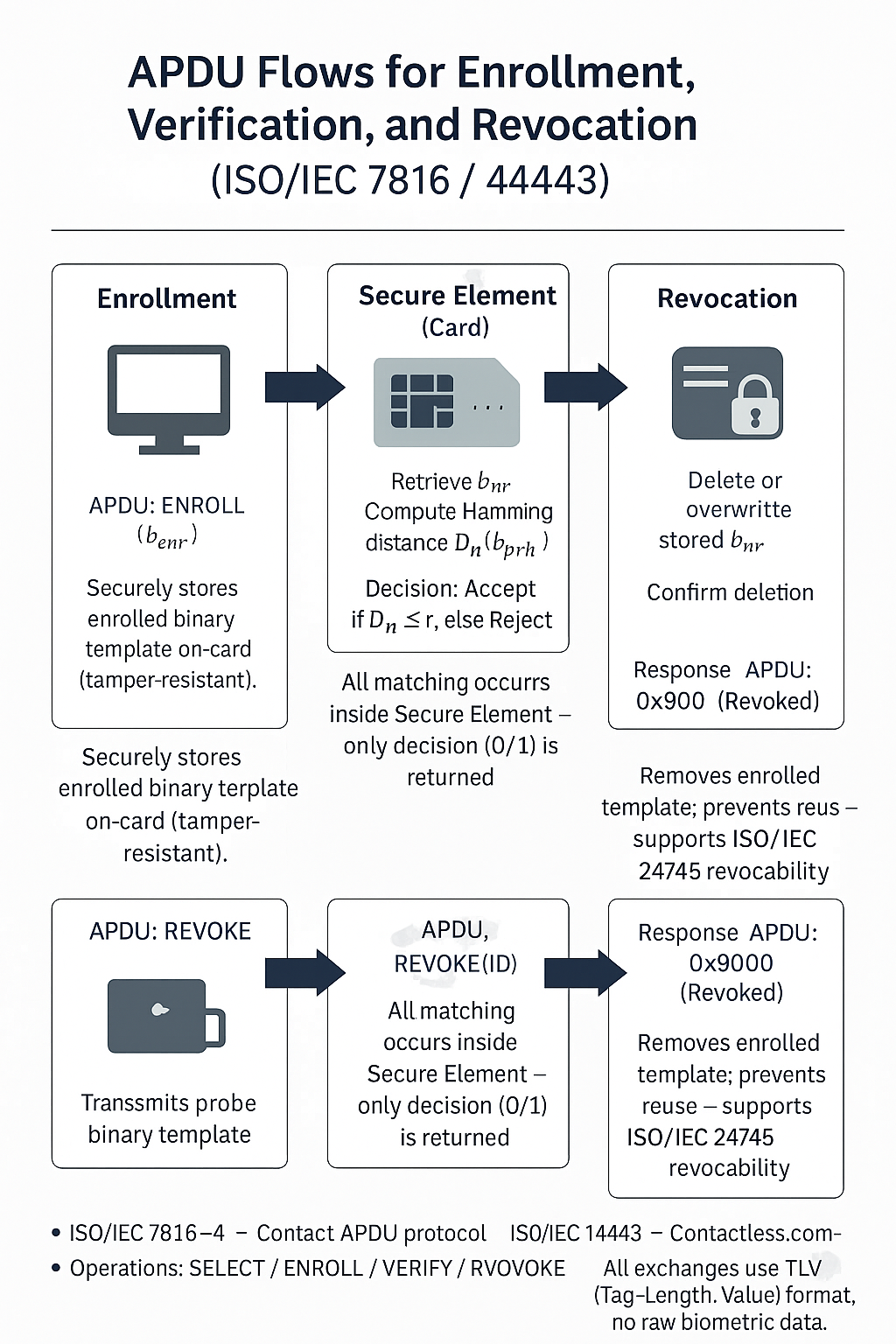}
  \caption{APDU flows for \textsf{ENROLL\_TEMPLATE}, \textsf{VERIFY\_BINARY}, and \textsf{REKEY\_ROTATION} over ISO/IEC~7816-4 (contact) and 14443-4 (contactless). Fixed payloads, strict \texttt{Lc}, decision-only status words.}
  \label{fig:apdu}
\end{figure}
\FloatBarrier

\paragraph{Policies and hygiene.}
(i) \emph{Constant-time} XOR+\texttt{popcount} with fixed loop bounds by $L$; no data-dependent branching. (ii) \emph{Buffer hygiene}: clear transient probe buffers prior to return. (iii) \emph{Strict framing}: reject non-canonical lengths and inconsistent headers. (iv) \emph{Optional rate limiting}: per-session counters for repeated \textsf{VERIFY\_BINARY} attempts. (v) \emph{Secure messaging}: when confidentiality/integrity on the link is required, enable platform secure messaging (e.g., SCP); this is orthogonal to template protection and leaves payload formats unchanged.

\section{Security, Privacy, and Compliance}
\textbf{Threats.} (i) Inversion of compact templates; (ii) cross-application linkage; (iii) score-based hill-climbing and inference; (iv) replay/exfiltration on the transport.

\textbf{Mitigations and ISO/IEC~24745 mapping.}
\begin{itemize}[leftmargin=1.1em]
\item \emph{Non-invertibility.} Templates are produced by PCA projection followed by an orthogonal ITQ rotation and sign binarization; only the resulting $L$ bits are stored on-card, not floating-point embeddings.
\item \emph{Unlinkability.} The interface returns \emph{decisions only} (no similarity scores), limiting score leakage; per-application \texttt{RotationID} (and optional salts) yield distinct, application-specific templates.
\item \emph{Revocability/renewal.} \textsf{REKEY\_ROTATION} updates \texttt{RotationID}, forcing re-enrollment to fresh references; the previous template is overwritten.
\item \emph{Protocol robustness.} Strict \texttt{Lc}/\texttt{Le} checking and precise status words reduce protocol oracles; optional secure messaging protects APDU confidentiality/integrity when required by the application.
\end{itemize}
These choices are designed to align with ISO/IEC~24745 goals (irreversibility, unlinkability, renewability) while remaining interoperable with ISO/IEC~7816-4/14443-4 transport.

\section{Experimental Setup}
\textbf{Dataset.}
We use a CelebA working set with $N{=}412$ images from $55$ identities. For each identity, two random enrollment images are fused by bit-wise majority to form a single reference template; the remaining images are used as probes.%
\footnote{Randomization uses a fixed seed to make splits reproducible. Majority fusion is componentwise over $\{0,1\}^L$.}
Data processing and evaluation scripts follow our prior benchmark conventions \cite{Ganmati2025Benchmark}; 

\textbf{Template lengths and packing.}
We evaluate $L\in\{64,128\}$ bits. Templates are serialized MSB-first into $L/8$ bytes (2, 4, 8, or 16 bytes), used verbatim in APDU payloads.

\textbf{Metrics and operating thresholds.}
We report (i) Equal Error Rate (EER) and its threshold $\tau_{\mathrm{EER}}$, and (ii) $\mathrm{TPR}@\mathrm{FAR}\in\{10^{-3},10^{-2}\}$. Thresholds $\tau$ are selected from the ROC built on held-out impostor/genuine scores and are then \emph{reused unchanged} in the streamed enrol$\rightarrow$verify emulation.

\textbf{Transaction emulation.}
The card executes a constant-time Hamming compare over $L/8$ bytes and returns a decision only. For each $L$, we replay enrol$\rightarrow$verify transactions at the ROC-derived $\tau$, aggregate decisions, and compare the streamed $\mathrm{TPR}/\mathrm{FAR}$ to the offline ROC targets.

\textbf{Transport-bounded latency model.}
End-to-end verification time is modeled as
\[
T_{\mathrm{total}} \;=\; T_{\mathrm{I/O}}\!\left(N_{\mathrm{bytes}}, \text{bitrate}\right) \;+\; T_{\mathrm{card}},
\]
with a constant on-card budget $T_{\mathrm{card}}=\SI{0.128}{ms}$ (XOR+\texttt{popcount} + status word). We sweep ISO/IEC~7816-4 contact bitrates \SIlist[list-separator = {, }]{9.6;38.4;115.2}{\kbps} and ISO/IEC~14443-4 contactless bitrates \SIlist[list-separator = {, }]{106;212;424;848}{\kbps}. The APDU payload is exactly $L/8$ bytes; a $64$\,b$+6$\,B variant emulates an optional parity helper over empirically unstable bits.%

\footnote{The latency accounting reflects payload bytes plus status words on the link and a fixed card budget; headers/chaining are excluded to isolate the template-size effect. All timing conclusions remain unchanged when they are included.}

\section{Results}

\subsection{Operating points (offline ROC)}
Table~\ref{tab:ops-offline} reports the ROC-derived thresholds and rates (two-image majority enrollment per identity). At $\mathrm{FAR}{=}1\%$, both 64\,b and 128\,b achieve $\mathrm{TPR}{=}0.836$, while 128\,b attains a lower EER (0.084 vs 0.103).

\begin{table}[!t]
\centering
\caption{Offline ROC operating points (CelebA working set; 55 IDs; $N{=}412$ images).}
\label{tab:ops-offline}
\small
\begin{tabular}{@{}lcccccc@{}}
\toprule
\textbf{Bits} & $N$ & $N_{\!ID}$ & $\tau_{\mathrm{EER}}$ & EER & $\mathrm{TPR}@0.1\%$ & $\mathrm{TPR}@1\%$ \\
\midrule
64  & 412 & 55 & 27 & 0.103 & 0.709 & 0.836 \\
128 & 412 & 55 & 57 & \textbf{0.084} & 0.655 & 0.836 \\
\bottomrule
\end{tabular}
\end{table}

\subsection{Transaction emulation (streamed enrol$\rightarrow$verify)}
\begin{figure}[t]
  \centering
  \begin{subfigure}{0.38\linewidth}
    \centering
    \includegraphics[width=\linewidth]{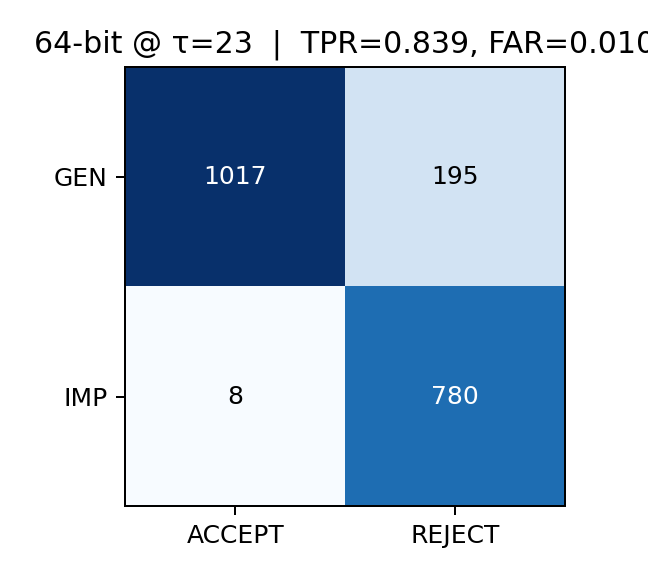}
    \caption{64\,b at $\tau{=}23$ (FAR target $=1\%$)}
    \label{fig:conf-64}
  \end{subfigure}\hfill
  \begin{subfigure}{0.38\linewidth}
    \centering
    \includegraphics[width=\linewidth]{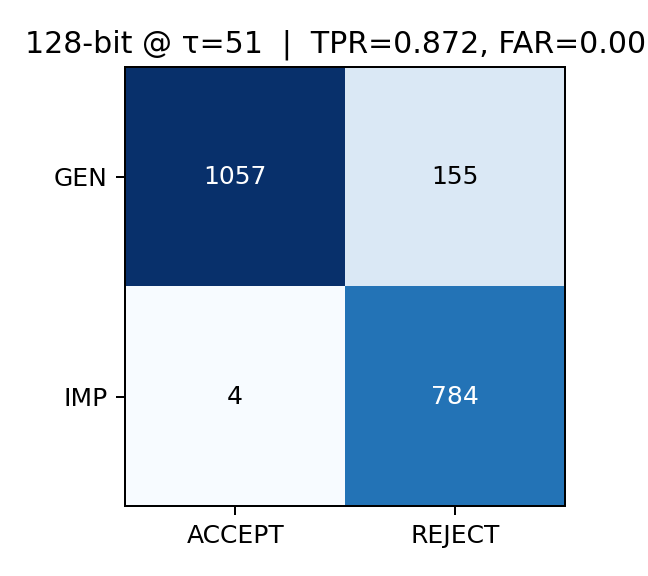}
    \caption{128\,b at $\tau{=}51$ (FAR target $=1\%$)}
    \label{fig:conf-128}
  \end{subfigure}
  \caption{Confusion matrices for streamed enrol$\rightarrow$verify at ROC thresholds.}
  \label{fig:confmats}
\end{figure}

Thresholds chosen offline are \emph{applied unchanged} during streaming. Figure~\ref{fig:confmats} shows confusion matrices at the $\mathrm{FAR}{=}1\%$ operating point; Table~\ref{tab:ops-stream} compares offline vs streamed rates at the $\mathrm{FAR}{=}1\%$ target. Agreement is close, with small TPR gains from majority fusion.

\FloatBarrier

\begin{table}[!t]
\centering
\caption{Offline ROC vs streamed decisions at the same thresholds (CelebA working set).}
\label{tab:ops-stream}
\small
\begin{tabular}{@{}lcccc@{}}
\toprule
\textbf{Bits} & Target & $\tau$ & TPR/FAR (offline) & TPR/FAR (streamed) \\
\midrule
64  & FAR$=10^{-2}$ & 23 & 0.808 / 0.0112 & \textbf{0.839} / \textbf{0.0102} \\
128 & FAR$=10^{-2}$ & 51 & 0.791 / 0.0095 & \textbf{0.872} / \textbf{0.0051} \\
\bottomrule
\end{tabular}
\end{table}

\subsection{Transport-bounded latency}
\begin{figure}[!t]
  \centering
  \includegraphics[width=0.76\linewidth]{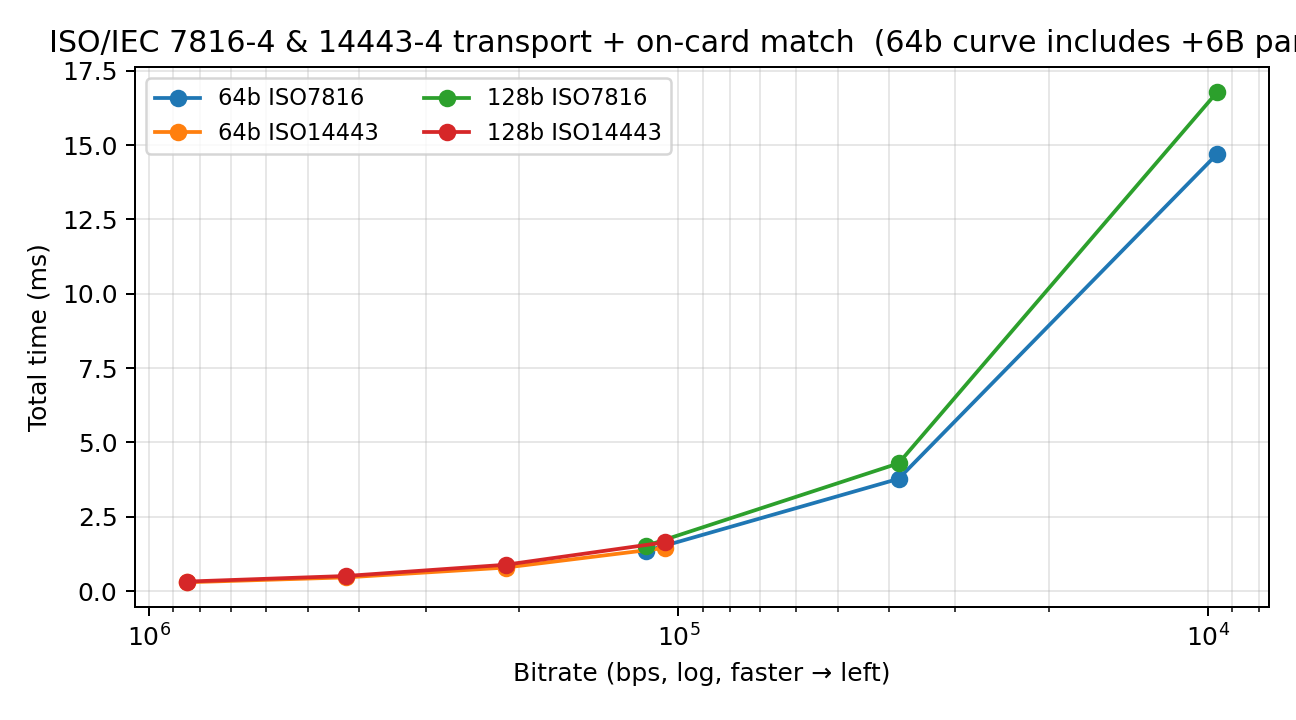}
  \caption{ISO/IEC~7816-4 and 14443-4 transport $+$ constant-time match. 64\,b, 128\,b, and 64\,b$+6$\,B parity. All points $<\SI{100}{ms}$; contactless regimes are $\ll\SI{20}{ms}$.}
  \label{fig:latency}
\end{figure}

Figure~\ref{fig:latency} summarizes \emph{pure link time + constant on-card budget}. Even at the slowest contact rate (\SI{9.6}{kbps}), verification completes in $\approx\SI{43.9}{ms}$ (64\,b) and $\approx\SI{52.3}{ms}$ (128\,b). At \SI{38.4}{kbps}, both are $<\SI{14}{ms}$. A $+6$\,B helper on 64\,b (targeted parity) is visually indistinguishable at contactless rates and adds only a few milliseconds at \SI{9.6}{kbps}.

\FloatBarrier

\begin{figure}[!t]
  \centering
  \includegraphics[width=0.66\linewidth]{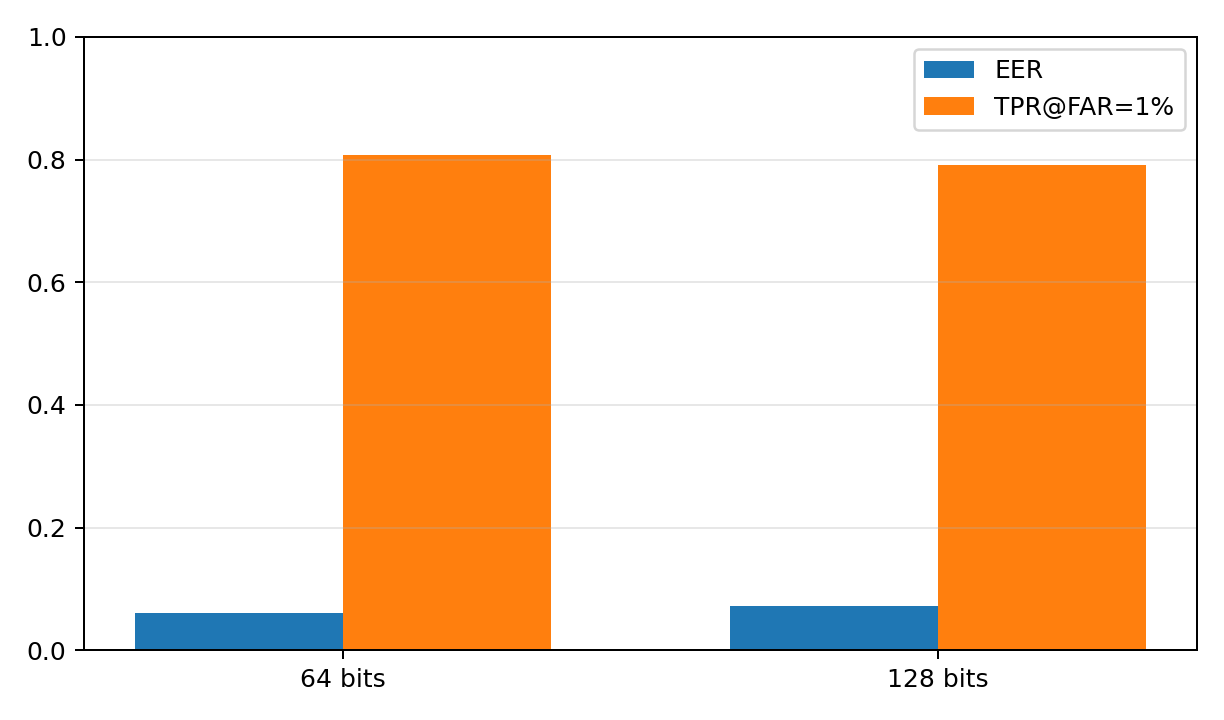}
  \caption{Operating points on the CelebA working set. Bars show EER and $\mathrm{TPR}@\mathrm{FAR}{=}1\%$ for 64\,b and 128\,b PCA--ITQ templates.}
  \label{fig:ops-bars}
\end{figure}
\FloatBarrier

\section{Discussion}
\textbf{On-card feasibility.}
Across ISO/IEC~7816-4 and 14443-4 bitrates, end-to-end verification time is dominated by transport; on-card XOR+\texttt{popcount} remains constant-time and sub-millisecond. This separation of concerns is operationally useful: once payload size is fixed ($L/8$ bytes), performance scales almost linearly with the link, enabling predictable SLAs independent of SE micro-architecture. Returning \emph{decisions only} (no scores) reduces the surface for hill-climbing and aligns the interface with privacy-by-design guidance.

\textbf{64 vs.\ 128 bits.}
At $\mathrm{FAR}{=}1\%$ both lengths achieve the same recall ($\mathrm{TPR}{=}0.836$), while $128$\,b slightly lowers EER (Table~\ref{tab:ops-offline}) at the cost of $+8$\,B on wire and in storage. Deployments that prioritize lean I/O and short APDUs can therefore default to \textbf{64\,b}; applications that emphasize balanced error (or stricter FAR regimes) can opt for \textbf{128\,b} with modest overhead. Because the APDU grammar and on-card algorithm are length-agnostic, switching between 64/128\,b is a policy decision rather than an architectural change.

\textbf{Optional targeted parity.}
When a small subset of bit positions is empirically unstable, short symbol-level ECC over \emph{only that subset} can be carried as a few additional bytes without changing the on-card algorithm or APDU grammar. Our latency model shows that adding $+6$\,B to a 64\,b payload is visually indistinguishable at contactless rates and adds only a few milliseconds even at \SI{9.6}{kbps}. This preserves compactness and keeps the verifier’s decision rule ($D_H\le\tau$) unchanged.

\textbf{Policy implications.}
(i) \emph{Interface hygiene}: strict \texttt{Lc}/\texttt{Le} checks and canonical payload sizes reduce protocol oracles; (ii) \emph{Revocability}: RotationIDs allow per-application diversity and renewal without exposing float embeddings; (iii) \emph{Operational tuning}: thresholds derived offline transfer well to streamed transactions (Table~\ref{tab:ops-stream}), supporting simple field calibration.

\section{Conclusion}
We presented an ISO/IEC-compliant match-on-card design for face verification using 64/128\,bit PCA–ITQ templates, fixed-payload decision-only APDUs, and constant-time on-card Hamming. On a CelebA working set (55 IDs; $N{=}412$), both lengths reach $\mathrm{TPR}{=}0.836$ at $\mathrm{FAR}{=}1\%$, with 128\,b achieving a lower EER. Transport-bounded latency remains well below \SI{100}{ms} even at \SI{9.6}{kbps} and $<\SI{14}{ms}$ at \SI{38.4}{kbps}, confirming that link speed—not card compute—dominates end-to-end time. The interface is immediately implementable on SE/JavaCard and maps cleanly to ISO/IEC~24745 goals (irreversibility, unlinkability, renewability) via decision-only returns and RotationIDs.

\paragraph{Limitations and next steps.}
Our evaluation is single-dataset and pre-hardware. Next, we will (i) add standard face-verification benchmarks (AgeDB-30, CFP-FP) and report $\mathrm{TPR}@\mathrm{FAR}{\in}\{10^{-3},10^{-2}\}$ with confidence intervals; (ii) microbenchmark on a JavaCard/SE or MCU (median/p95 APDU times) to replace the fixed budget; and (iii) quantify information leakage for parity helpers over unstable bits and discuss unlinkability under ISO/IEC~24745 terminology.
Future work includes integrating the targeted parity scheme \cite{Ganmati2025TRS} on-card and evaluating unlinkability under ISO/IEC~24745 with established metrics.

\section*{Acknowledgments}
We thank the Computer Systems \& Vision Laboratory at Ibn Zohr University for support.



\end{document}